\documentclass[aps,prl,twocolumn,superscriptaddress,showpacs,floatfix]{revtex4}
\usepackage{epsfig}
\newcommand{\be}{\begin{equation}}
\newcommand{\ee}{\end{equation}}
\newcommand{\bc}{\begin{center}}
\newcommand{\ec}{\end{center}}
\newcommand{\bi}{\begin{itemize}}
\newcommand{\ei}{\end{itemize}}
\newcommand{\ba}{\begin{eqnarray}}
\newcommand{\ea}{\end{eqnarray}}

\newcommand{\ignore}[1]{}

\begin{document}
\draft
\title{Scale-free brain functional networks}.

\author{Victor M. Egu\'{\i}luz} \affiliation{Instituto
Mediterr\'aneo de Estudios Avanzados, IMEDEA (CSIC-UIB), E07122
Palma de Mallorca, Spain}

\author{Dante R. Chialvo}
\affiliation{Department of Physiology, Northwestern University,
Chicago, Illinois, 60611}

\author{Guillermo A. Cecchi}
\affiliation{IBM T.J. Watson Research Center, 1101 Kitchawan Rd.,
Yorktown Heights, NY 10598}

\author{Marwan Baliki}
\affiliation{Department of Physiology, Northwestern University,
Chicago, Illinois, 60611}
\author{A. Vania Apkarian}
\affiliation{Department of Physiology, Northwestern University,
Chicago, Illinois, 60611}

\date{\today}

\begin{abstract}

Functional magnetic resonance imaging (fMRI) is used to extract
{\em functional networks} connecting correlated human brain sites.
Analysis of the resulting networks in different tasks shows that:
(a) the distribution of functional connections, and the
probability of finding a link vs. distance are both scale-free,
(b) the characteristic path length is small and comparable with
those of equivalent random networks, and (c) the clustering
coefficient is orders of magnitude larger than those of equivalent
random networks. All these properties, typical of scale-free small
world networks, reflect important functional information about
brain states.

\end{abstract}
\maketitle

Recent work has shown that disparate systems can be described as
complex networks, that is assemblies of nodes and links with
nontrivial topological properties, examples of which include
technological, biological and social systems \cite{albert2002}.
The brain is inherently a dynamic system, in which the traffic
between regions, during behavior or even at rest, creates and
reshapes continuously complex functional networks of correlated
dynamics. An important goal in neuroscience is to understand these
spatio-temporal patterns of brain activity. This Letter proposes a
method to extract functional networks, as revealed by fMRI in
humans, and analyze them in the context of the current
understanding of complex networks (for reviews see
\cite{albert2002,strogatz2001,watts1998}).

Figure 1 shows how underlying functional networks are exposed
during any given task. In these experiments, at each time step
(typically 400 spaced 2.5 sec.), magnetic resonance brain activity
is measured in $36\times64\times64$ brain sites (so-called
``voxels'' of dimension $3\times 3.475\times 3.475~$mm$^3$). The
activity of voxel $x$ at time $t$ is denoted as $V(x,t)$. We
define that two voxels are {\sl functionally connected} if their
temporal correlation exceeds a positive pre-determined value
$r_c$, regardless of their {\sl anatomical connectivity}
\cite{dodel2002,note}. Specifically, we calculate the linear
correlation coefficient between any pair of voxels, $x_1$ and
$x_2$, as: \be r(x_1, x_2)= \frac{\langle V(x_1,t) V(x_2,t)\rangle
- \langle V(x_1,t)\rangle \langle V(x_2,t) \rangle} {\sigma
(V(x_1))\sigma (V(x_2))} ~, \ee where $\sigma^2(V(x)) = \langle
V(x,t)^2\rangle - \langle V(x,t)\rangle^2$, and $\langle \cdot
\rangle$ represents temporal averages.
\begin{figure}[tb]
\centering \psfig{figure=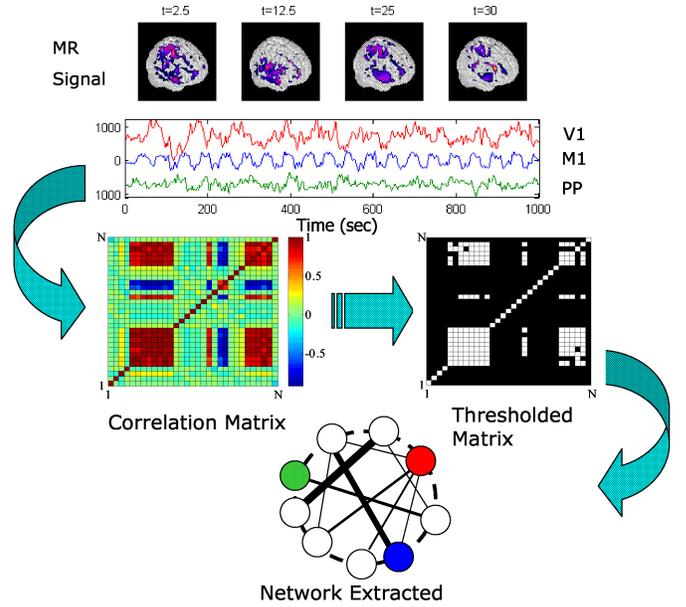,width=3.5
truein,clip=true,angle=0} \caption{\footnotesize{(Color online)
Methodology used to extract functional networks from the signals.
The correlation matrix is calculated and then used to define the
network among the highest correlated nodes. Top four images
represent snapshots of activity and the three traces correspond to
selected voxels from visual (V1), motor (M1) and posterio-parietal
(PP) cortices.}}
\end{figure}

Figure 2 shows the degree distributions of networks extracted
using this method. The data was collected while the subject was
opposing fingers 1 and 2 during 10 seconds, and then resting
during 10 sec. We find a skewed distribution of links with a tail
approaching a distribution $p(k) \sim k^{-\gamma}$, with $\gamma$
around 2. This power law is more evident for networks constructed
with higher thresholds $r_c$ (more correlated conditions). For
decreasing $r_c$, a maximum appears which shifts to the right.
Despite changes in parameters, networks remain clearly defined
indicating that the main conclusions are robust with respect to
the selection of parameters. The small inset in Fig.~2 shows the
distribution of links of a network constructed from the randomly
shuffled (in time) voxels' signal. This network  displays a
Gaussian degree distribution in which the mean and width depend on
$r_c$. The largest values of the correlation thresholds used to
construct the random networks are usually extremely low ($r_c \sim
0.1$) compared to that used to define the functional networks
($r_c \sim 0.7$). Our data was also compared with values from a
randomly re-wired network, where nodes keep their degree by
permuting links (i.e., the link connecting nodes $i,j$ is permuted
with that connecting nodes $k,l$) \cite{maslov2003} (see below).
In this control the degree of each  node is maintained but all
other correlations (including clustering) are destroyed.

\begin{figure}[tb]
\centering
 \psfig{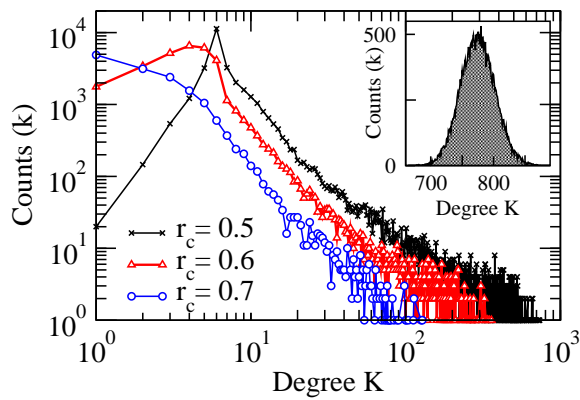}
 \caption{\footnotesize{(Color online)
Degree distribution for three values of the correlation threshold.
The inset depicts the degree distribution for an equivalent
randomly connected network.}}
\end{figure}

To test the generality of these findings the same analysis was
performed in 7 subjects across 3 task conditions. During data
acquisition \cite{methods} subjects perform on-off finger tapping
with three different protocols. In one case they are instructed
verbally to start and stop tapping, in the other one the
start/stop cue is a small green/red dot in a video screen, and in
the last one the start/stop cue is the entire screen turning green
or red. The results are very robust across subjects and task
conditions. In particular, the average of degree distribution (see
Fig.~3) shows a clear power law scaling decaying as $p(k) \sim
k^{-\gamma}$, with an exponent close to 2. Although a precise
fitting is arguably difficult, we find that for $r_c$=0.6 $\gamma$
= 2, for $r_c$=0.7 is 2.1 and for $r_c$=0.8 is 2.2. This power
law, indicating that the functional networks are scale-free,
implies that there is always a small but finite number of brain
sites having broad ``access'' to most other brain regions. Those
well connected nodes are comparatively much more numerous in these
networks than in a randomly connected network.

As shown in the bottom panel of Fig.~3 the average probability of
finding a link between two nodes, separated at least by a distance
$\Delta$, also decays as a power law. The significance of the
scaling with distance is unclear because of the well known
extensive cortex folding, which makes linear distance a dubious
parameter.

\begin{figure}[tb]
\centering \psfig{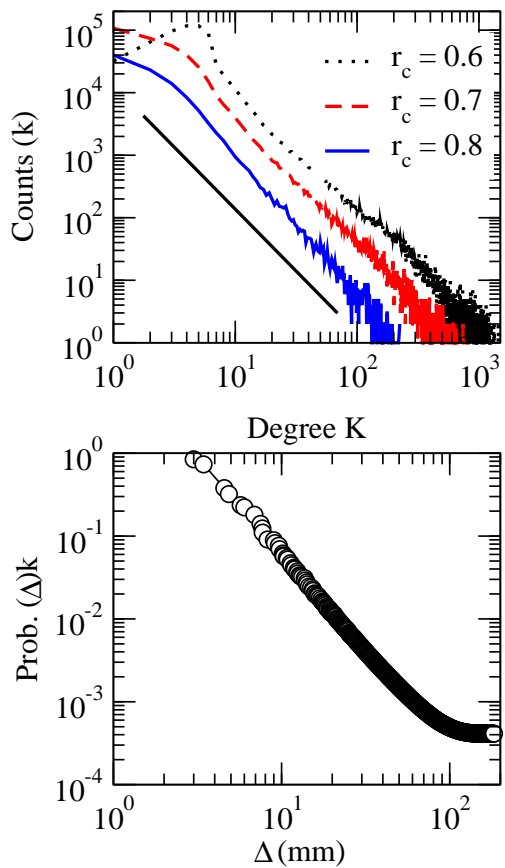} \caption{\footnotesize{(Color online)
Average scaling taken from 22 networks extracted from seven
subjects. Top Panel: Average degree distribution. The straight
line illustrates a decay of $k^{-2}$. Bottom panel: Average
probability of finding a link between two nodes separated by a
distance larger than $\Delta$ (using $r{_c}=0.6$).}}
\end{figure}

The scale-free character remains unaltered even for tasks engaging
different brain regions. This is already implicit in the
aggregated data of Fig.~3 (top panel), but we further corroborated
this feature by analyzing two radically different brain states:
listening to music and finger tapping. As shown in Fig.~4,
although the topographic distribution of the functional networks
is very different for the two tasks, they have similar scaling
behavior. For comparison, the standard activation map derived with
the generalized linear model \cite{friston2002} is also shown.

Now we turn to describe statistical properties of these networks:
path length and clustering. The path length ($L$) between two
voxels is the minimum number of links necessary to connect both
voxels. Clustering ($C$) is the fraction of connections between
the topological neighbors of a voxel with respect to the maximum
possible. If voxel $i$ has degree $k_i$, then the maximum number
of links between the $k_i$ neighbors is $k_i(k_i-1)/2$. Thus, if
$E_i$ is the number of links connecting the neighbors then the
clustering of voxel $i$, $C_i = 2E_i/ k_i(k_i-1)$. The average
clustering of a network is given by $C = 1/N\sum_i C_i$, where $N$
is the number of voxels. Clustering was analyzed also with respect
to degree. The average clustering over voxels with the same degree
$C(k) = 1/N_k \sum_{j=\{i|k_i = k\}} C_j $, where the sum runs
over the $N_k$ voxels with degree $k$.

\begin{figure}[tb]
\centering
\psfig{figure=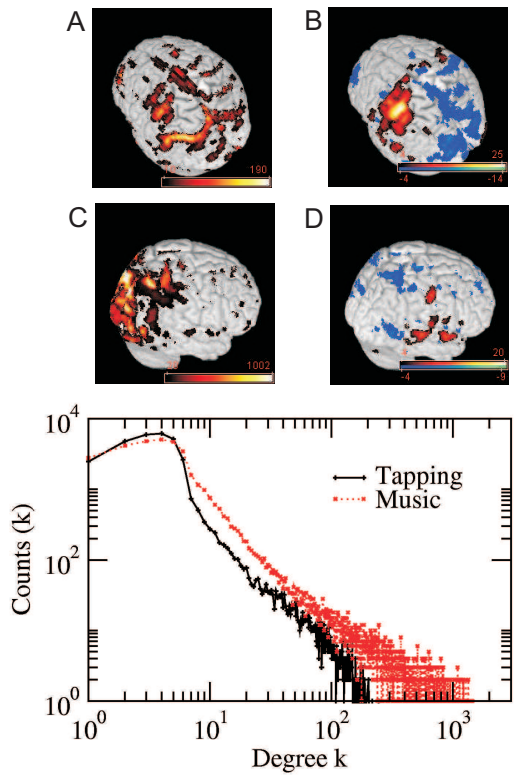,width=3truein,clip=true,angle=0}
\caption{\footnotesize{(Color online) Comparison for two tasks:
Panels A and B correspond to a finger tapping task while  C and D
to listening to music analyzed with our method or the standard
fMRI linear model. Colors in pictures of panels A and C code the
number of links detected with our method, and those in panels B
and D the activation map build with standard model
\cite{friston2002}. The link distributions (lower panel) show that
the networks for both tasks are scale free.}}
\end{figure}

Table 1 summarizes the results for the networks analyzed showing
the average values (n = 22 datasets) for each threshold ($r_c$,
first column) used to construct the networks. Listed are $N$, $C$,
$L$, the average degree $\langle k \rangle$, and $\gamma$. The
clustering ($C_{rand}$) and path length ($L_{rand}$) values of an
equivalent random network are also included for comparison. Note
that as the threshold $r_c$ increases the total number of nodes
$N$ decreases substantially, resulting by definition in more
correlated networks. As a result, the number of nodes with at
least one link decreases, and consequently the $\langle k \rangle$
value decreases as well. In all cases, the coefficient $C$ remains
four orders of magnitude larger than $C_{rand}$. Networks
randomized using the rewiring described by Maslov et al.
\cite{maslov2003} also have clustering significantly smaller than
the raw data (the order of $10^{-2}$). This feature, together with
the similarity of path length of the original nets and their
randomized controls ($L$ and $L_{rand}$), is indicative of a
small-world structure \cite{strogatz2001,watts1998}. This property
is robust as it does not depend on parameter $r_c$.

\begin{table}
 \centering
\begin{tabular}{|c|c|c|c|c|c|c|c|}
 \hline
 $r_{c}$& $N$ & $C$ & $L$ & $\langle k \rangle$ &$\gamma$& $C_{rand}$& $L_{rand}$ \\
 \hline
0.6 & 31503 & 0.14 & 11.4 & 13.41 & 2.0 & 4.3$\times 10^{-4}$ & 3.9 \\
0.7 & 17174 & 0.13 & 12.9 & 6.29 & 2.1& 3.7$\times 10^{-4}$&5.3 \\
0.8 & 4891 & 0.15 & 6. & 4.12 & 2.2& 8.9$\times 10^{-4}$& 6.0 \\
 \hline
 \end{tabular}\\
 \caption{Average statistical properties of the brain functional networks.}
\end{table}

\begin{table}
 \centering
\begin{tabular}{|c|c|c|c|c|c|c|c|}
 \hline
 $Network$& $N$ & $C$ & $L$ & $\langle k \rangle$ &$\gamma$& $C_{rand}$& $L_{rand}$ \\
 \hline
C. Elegans & 282 & 0.28 & 2.65 & 7.68 & NA & 0.025 & 2.1 \\
Macaque VC & 32 & 0.55 & 1.77 & 9.85 & NA & 0.318 &1.5 \\
Cat Cortex & 65 & 0.54 & 1.87 & 17.48 & NA & 0.273 &1.4\\
 \hline
 \end{tabular}\\
 \caption{Previously reported statistics of
 relatively smaller networks. None of these networks is scale-free.}
\end{table}

To our knowledge, this is the first report on the topological
structure of a large scale brain network. Previous studies
employing these statistical analyses have been limited to the
small data sets of C. Elegans \cite{watts1998}, and two
neuro-anatomical databases \cite{sporns2000,hilgetag2000}, the
macaque visual cortex \cite{felleman1991} and the cat cortex
\cite{scannel1999}(see Table 2). These studies did not demonstrate
scale-free features. Comparison with the previous two reports
indicate the following: although clustering in the present study
is smaller in absolute value, it is still orders of magnitude
larger than the random case ($10^{-1}$ vs. $10^{-4}$), while in
the previous reports the clustering of the experimental data was
just one order of magnitude larger than the randomized controls in
the best case. Interestingly, the average connectivity $\langle k
\rangle$ in all cases is of the same order, despite the huge
differences in networks' origins and sizes. Accordingly, this
consistency may reflect some constraint(s) inherent to network
construction. These quantitative features show that the human
brain network examined here has small world properties, a finding
that was previously postulated \cite{strogatz2001,watts1998}.

Figure 5 illustrates the dependence of two important features upon
a voxel's degree. The first is clustering, found in many cases to
scale as $C(k) \sim k^{-\alpha}$, an indication of hierarchical
organization \cite{ravasz2003,jeong2001}. We see, instead, a
relative independence of clustering from degree. The second
feature is that a highly connected node tends to connect with
other well connected nodes. As shown in the bottom panel of Figure
5, there is a positive correlation between the degrees of adjacent
vertices. This correlation, also called assortative mixing, is not
typical of biological networks, but rather is distinctive of
social networks \cite{newman2002}. Transitivity in correlations
contributes to increase artifactually the clustering coefficient,
using partial directed coherence or Granger causality
\cite{baccala} in the future should clarify this.

\begin{figure}[tb]
\centering \psfig{figure=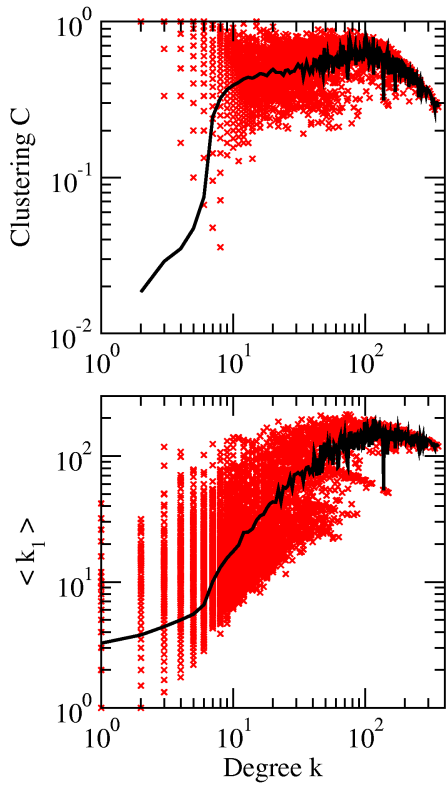,width=2.5
truein,clip=true,angle=0} \caption{\footnotesize{(Color online)
Top Panel: Plot of clustering vs. degree. Bottom panel: Plot of a
neighboring node's degree vs. degree illustrates the assortative
feature. Symbols represents individual data and continuous lines
the average values for nodes with the same degree. (Same subject
shown in Fig. 2, with $r_c$=0.6).}}
\end{figure}

In summary, we report statistical measures showing that the
functional correlations of the human brain form a scale-free
network with small world properties and assortative mixing. While
some of these properties have been informally discussed, this work
is the first quantitative description of these large-scale
topological properties, as well as the first report of an
assortative biological network. The scaling laws demonstrated here
are robust across parameters (Fig. 1), subjects (Fig. 3), and task
conditions (Fig. 4), suggesting they are invariant properties of
an underlying dynamical network. The present results complement
the extensive work done in the context of brain functional and
effective connectivity \cite{biswal1995,friston2002}. The present
approach has additional important implications. Namely, these
studies can be extended to cases in which standard fMRI techniques
cannot be used for lack of subject cooperation, (e.g., Alzheimer's
patients). Because scale-free complex networks are known to show
resistance to failure, facility of synchronization, and fast
signal processing \cite{lago2000}, it would be important to see
whether brain networks scaling properties are altered under
various pathologies. In that regard, techniques for investigation
of communities' structures \cite{radicchi2003} should be useful to
analyze these aspects. Work on models \cite{caldarelli2002}, is
needed to further clarify specific origins of the scaling laws.
Overall, the network properties uncovered here, offer a novel
window to investigate the dynamics of brain states particularly in
cases of dysfunction.

Work supported by MCyT of Spain (Projects CONOCE2, BFM2002-12792-E
and FIS2004-05073-C04-03) and NIH NINDS of USA (Grants 42660 and
35115). DRC is grateful for the hospitality and support of the
Universitat de les Illes Balears, Mallorca, Spain.


\begin{thebibliography}{999}
\bibitem{note}Networks can be defined using negative correlations,
we restrict ourselves to the positive case only for simplicity.
\end{thebibliography}
\end{document}